\documentclass[aps,amsmath,amssymb,groupedaddress,twocolumn]{revtex4}
\usepackage{graphicx}

\newcommand{\nc}{\newcommand}
\nc{\beq}{\begin{equation}}
\nc{\eeq}{\end{equation}}
\nc{\beqa}{\begin{eqnarray}}
\nc{\eeqa}{\end{eqnarray}}

\def\gsim{\mathrel{\rlap{\lower4pt\hbox{\hskip1pt$\sim$}}
    \raise1pt\hbox{$>$}}}       %greater than or approx. symbol

\begin{document}

\title{Black hole entropy, curved space and monsters}

\author{Stephen~D.~H.~Hsu} \email{hsu@uoregon.edu}
\author{David Reeb} \email{dreeb@uoregon.edu}
\affiliation{Institute of Theoretical Science \\ University of Oregon,
Eugene, OR 97403}

\begin{abstract}
We investigate the microscopic origin of black hole entropy, in
particular the gap between the maximum entropy of ordinary matter and
that of black holes. Using curved space, we construct configurations
with entropy greater than the area $A$ of a black hole of equal mass. These
configurations have pathological properties and we refer to them as
{\it monsters}. When monsters are excluded we recover the entropy
bound on ordinary matter $S < A^{3/4}$.  This bound implies that
essentially all of the microstates of a semiclassical black hole are
associated with the growth of a slightly smaller black hole which
absorbs some additional energy. Our results suggest that the area
entropy of black holes is the logarithm of the number of distinct ways
in which one can form the black hole from ordinary matter {\it and}
smaller black holes, but only after the exclusion of monster states.
\end{abstract}

%\pacs{}

\maketitle

\date{today}

\bigskip

Black holes radiate \cite{Hawking} and have entropy $S = A/4$, where
$A$ is the surface area in Planck units \cite{Bekenstein}. The nature
of this entropy is one of the great mysteries of modern physics,
especially due to its non-extensive nature: it scales as the area of
the black hole in Planck units, rather than its volume. This peculiar
property has led to the holographic conjecture
\cite{HoltHooft,HolSusskind} proposing that the number of degrees of
freedom in any region of our universe grows only as the area of its
boundary. The AdS/CFT correspondence \cite{AdSCFTreview} is an
explicit realization of holography.

The entropy of a thermodynamic system is the logarithm of the number
of available microstates of the system, subject to some macroscopic
constraints such as fixed total energy. For a black hole, this means
all possible internal states with fixed total mass, charge and angular
momentum. In certain string theory black holes, these states have been
counted explicitly \cite{SV,MSW}. As a proxy for counting microstates,
we might instead count the number of distinct ways of forming a black
hole \cite{ZT}, since each distinct pre-configuration presumably
corresponds to a unique microstate.

It is easy to see that gravitational collapse limits the entropy of
physical systems. Information (entropy) requires energy, while
gravitational collapse (formation of a horizon or black hole)
restricts the amount of energy allowed in a finite region
\cite{Hsu}. 't Hooft \cite{HoltHooft} showed that if one excludes
configurations whose energies are so large that they will inevitably
undergo gravitational collapse, one obtains $S <
A^{3/4}$.  To deduce this result, 't Hooft replaces the system under
study with a thermal one. This is justified because, in the large
volume limit, the entropy of a system with constant total energy $E$
(i.e., the logarithm of the phase space volume of a microcanonical
ensemble) is given to high accuracy by that of a canonical ensemble
whose temperature has been adjusted so that the average energies of
the two ensembles are the same. 

Given a thermal region of radius $R$ and temperature $T$, we have $S
\sim T^3 R^3$ and $E \sim T^4 R^3$. Requiring $E < R$ (using the hoop
conjecture \cite{hoop,bh}) then implies $T < R^{-1/2}$ and $S <
R^{3/2} \sim A^{3/4}$. We stress that the use of temperature here is
just a calculational trick -- the result can also be obtained by
directly computing the volume of phase space on a surface of fixed
energy, as limited by the collapse condition.

In \cite{Buniy:2005au}, it 
was shown that imposing the condition ${\rm Tr} [ \, \rho H \,] < R$
on a density matrix $\rho$ implies a similar bound $S_{\rm vN} <
A^{3/4}$ on the von Neumann entropy $S_{\rm vN} = - {\rm Tr} \, \rho \ln
\rho$. For $\rho$ a pure state the result reduces to the previous
Hilbert space counting.  

We note that these bounds are more restrictive than the bound obtained
from black hole entropy: $S < A / 4$. Is there a gap between the
maximum entropy of matter configurations and that of black holes? If
so, it would imply that microstates of a large black hole are
overwhelmingly dominated by those originating from a slightly smaller
black hole \cite{fn}. The $\exp (A^{3/4})$ matter configurations without
horizons would be negligible compared to the $\exp (A/4 \, - \, \delta)$
slightly smaller black holes that might, upon the addition of a small
amount of energy, have formed a given black hole of area $A$.

\bigskip
{\bf Packing it in}
\bigskip

't Hooft's calculation described above is done in flat space, taking spatial 
volume to be proportional to $R^3$. The only appearance of general
relativity is in the hoop conjecture. We now show that the $A^{3/4}$
and $A$ entropy bounds can be exceeded by matter configurations in
curved space, in effect by changing the relationship between internal
volume and surface area. A technical remark: in a general curved spacetime the ``size'' or ``area" of a region is difficult to define in a coordinate-independent way. However, in the case of spherical symmetry, which we assume here, these issues do not arise  \cite{Wald:1999vt}. Moreover, what we are primarily interested in is the entropy of our configuration relative to the area of a black hole of equal mass, into which it will evolve.

We consider spherically symmetric, but not necessarily static,
distributions of matter, using standard coordinates
\begin{equation}
ds^2=-g_{tt}(r,t)dt^2+ g_{rr}(r,t) dr^2+r^2 d \Omega^2~~.
\end{equation}
Further, we define
\begin{equation}
\label{eps}
\epsilon(r) = 1 - \frac{2 M(r)}{r}~~,
\end{equation}
with (``energy within radius r'')
\begin{equation}
\label{M}
M(r) = 4 \pi \int_0^r dr' \, {r'\,}^2 \rho(r')~~,
\end{equation}
where $\rho(r)=\rho(r,t_0)$ is the proper energy density (i.e., as
seen by a stationary observer at $r$) on the initial time slice
$t=t_0$. Then, assuming the matter to be initially at rest w.r.t.~our
$(r,\theta,\phi)$ coordinates, the metric on that slice is fully
determined by \cite{pavelle}
\begin{equation}
g_{rr} (r,t_0)=\epsilon(r)^{-1}~~.
\end{equation}
The total mass (or ADM energy) is simply $M\equiv M(R)$ if $R$ is the
radius of the distribution. The total entropy is obtained as
follows. First, assume the existence of a covariantly conserved
entropy current $j^{\mu}$, i.e.~$j^{\mu}_{~\, ;\mu}=0$. (If entropy is
not conserved, the second law requires that it always increases, which
means our result is still a lower bound for any black hole produced.)
From the Stokes theorem we have that
\begin{equation}
\label{ss}
S_{\, \Sigma} ~=~ \int_\Sigma d^3x \sqrt{\gamma} \, s ~=~ {\rm constant}~~,
\end{equation}
where the integral is taken over a constant time slice $\Sigma$ with
induced metric $\gamma$ and unit normal $n^{\mu} \sim
(\partial_t)^{\mu}$, and $s=-j^{\mu}n_{\mu}$ is the proper entropy
density (as seen by a stationary inertial observer). In our
coordinates, $s(r)=j^0(r,t_0) g_{tt} (r,t_0)^{1/2}$ and the total
entropy of the initial configuration on the time 
slice $t = t_0$ is given by
\begin{equation}
\label{S}
S = 4 \pi \int_0^R dr \, r^2 \epsilon(r)^{-1/2} s(r)~~.
\end{equation}
For related discussion, see \cite{weinberg}.

Note that the {\it proper} mass of our object is 
\begin{equation}
M_p = 4 \pi \int_0^R dr \, r^2 \epsilon(r)^{-1/2} \rho(r)~~,
\label{pmass}
\end{equation}
and the difference between $M$ and $M_p$ is the negative binding
energy. As discussed below, the ratio $M / M_p$ can be made as small
as desired for large $R$ \cite{Hsuzero}.

To ensure that our object is not already a black hole, we require
$\epsilon (r) > 0$ at all $r$. Subject to this constraint, we attempt
to maximize $S$. The resulting entropy is a lower bound on the
entropy of a black hole of radius $R$. Here we of
course refer to the internal state of the hole; from the outside they
are identical.

We take $s(r) \sim \rho(r)^{3/4}$, which is appropriate for
relativistic matter. For thermal matter we would have $s \sim T(r)^3$
while $\rho \sim T(r)^4$, where $T(r)$ is the temperature at radius
$r$. Note we do not assume our configuration is in thermal
equilibrium; temperature is used here to count the number of initial
configurations with the desired energy density profile $\rho (r)$, as
in the case of the flat space calculation. Another possibility is a
relativistic Fermi gas, in which the energy scale is determined by the
Fermi momentum.

The difference between the curved and flat space cases is due entirely
to the factor of $\epsilon(r)^{-1/2}$ in integrals like Eq.~(\ref{S}) and
Eq.~(\ref{pmass}). Consequently, the flat space bound of $S < A^{3/4}$ can
only be exceeded for configurations in which $\epsilon (r)$ is close
to zero  (equivalently, $2M(r) \approx r$) 
in a subregion containing significant entropy and energy
density. In fact, for any configuration in which $\epsilon (r) >
\epsilon_0$ for all $r$, one can easily deduce that $S <
\epsilon_0^{-1/2} A^{3/4}$, since by removing $\epsilon (r)$ from the
integral in Eq.~(\ref{S}) one is left with the flat space entropy.

Some explicit examples are given below, in which curved space allows
violation of both the $A^{3/4}$ and $A$ entropy bounds. (See \cite{MS}
for a discussion of highly
entropic objects and their effect on black hole thermodynamics.) 
Subsequently, we will show that
configurations with significant energy density in regions with
$\epsilon(r) \approx 0$ have
pathological properties, and we will refer to them as monsters.

\bigskip
{\bf Example 1: blob of matter}
\bigskip

As a simple example, consider an object with a small core of radius
$r_0$ and mass $M_0$ and density profile
\begin{equation}
\rho(r) = \rho_0 \left( \frac{r_0}{r} \right)^2 ~~~~ (r_0 < r < R)~~.
\end{equation}
Then
\begin{equation}
M(r) = M_0 + 4 \pi \rho_0 r_0^2 (r - r_0)~~.
\end{equation}
We choose $8 \pi \rho_0 r_0^2 = 1$ so that
\begin{equation}
\epsilon (r) = \epsilon_0 \left( \frac{r_0}{r} \right)~~,
\end{equation}
where $\epsilon_0 = 1 - 2M_0 / r_0$.

From Eq.~(\ref{S}), the total entropy of this object is (neglecting
the small core region $r < r_0$)
\begin{equation}
S ~\sim~ 4 \pi \int^R_{r_0} dr \, r^2 \left( \frac{r}{r_0
\epsilon_0} \right)^{1/2} \rho^{3/4} ~\sim~ {\rho_0^{3/4} r_0 \over
\sqrt{\epsilon_0}} \, R^2 ~~.
\label{S1}
\end{equation}
Note that area scaling has been achieved. The overall entropy $S$
can be made as large as desired by taking $\epsilon_0$ small. We can
also obtain faster than area scaling by taking $\epsilon (r)$ to
approach zero faster than $1/r$.

\bigskip
{\bf Example 2: thin shell}
\bigskip

Consider a thin shell of material with $R < r < R+d$. 

We first consider the class of models $M(r) = (R+d) z^n / 2$, where $z
= (r-R)/d$ and $n > 0$. In these models the mass of the shell increases
smoothly to the maximum possible value as $r$ approaches $R+d$.

We write the energy density $\rho (r)$ as
\begin{equation}
\rho(r) = \frac{M'(r)}{4 \pi r^2}~~,
\end{equation}
where prime denotes differentiation with respect to $r$. Then,
the entropy of the shell is given by
\begin{equation}
S = (4 \pi)^{1/4} \int_R^{R+d} dr \, r^{1/2}
\frac{M'(r)^{3/4}}{\epsilon(r)^{1/2}}~~.
\end{equation}
Taking $d$ much less than $R$,
\begin{equation}
S \sim R^{5/4} d^{1/4} \int_0^{1} dz \,
\frac{ n^{3/4} z^{3(n-1)/4} } { (1-z^n)^{1/2}}~~.
\end{equation}
The $z$ integral is convergent for $n > 0$, so the entropy scaling is
given by $S \sim R^{5/4} d^{1/4}$ or at most $S \sim A^{3/4}$ if $d$
scales at most as $R$.

However, we can also construct thin shell configurations with unbounded
entropy. For example, take $\epsilon (r)$ to decrease rapidly to some
$\epsilon_0$ between $R$ and $R_1$:
\begin{equation}
M(r) = \frac{r - R}{2(R_1 - R)} R_1 (1 - \epsilon_0) ~~~~ (R < r < R_1)~,
\end{equation}
and then hold $\epsilon (r) = \epsilon_0$ for $r > R_1$:
\begin{equation}
M(r) = \frac{r}{2} (1 - \epsilon_0) ~~~~ (R_1 < r < R+d)~.
\end{equation}
The entropy in the region $R_1 < r < R+d$ can be made as large as
desired by taking $\epsilon_0$ sufficiently small.

\bigskip

As demonstrated, curved space configurations can have greater entropy
than their flat space counterparts of the same mass or size. 
This is because of their small $\epsilon(r)$: the configurations have proper surface area $A \sim M^2$, but have internal proper volume much larger than $A^{3/2}$.
Equivalently, they have very large proper mass $M_p$ relative
to mass $M$. It is easy to see that the ratio $M / M_p$
%\begin{equation}
%{M \over M_p} = {\int^R dr \, r^2 \rho(r) \over \int^R dr \, r^2
%\epsilon(r)^{-1/2} \rho(r)}  ~~
%\end{equation}
can be made as small as desired if $\epsilon(r)$ approaches zero for
large $r$. The large negative gravitational binding energy allows us
to pack substantially more proper mass into the region than suggested
by a flat space analysis. 

Regarding coordinate invariance of our results, we note that the total entropy $S$ of the initial configuration on the time slice 
$t = t_0$ is, by construction in (\ref{ss}), coordinate-invariant. Also, 
the area $A$ of a black hole formed by one of our configurations (by construction, on the verge of collapse) is simply a function of the ADM mass $M$, which is invariant. Of physical interest here is the entropy of our configuration compared to the area $A$ of a black hole of equal mass.

Without a constraint on how close $\epsilon (r)$ can get to zero, $S$
can be made arbitrarily large. Invoking quantum effects, one might
require that a Planck length uncertainty \cite{CGH} in the proper
radial distance not cause horizon formation, i.e.~that $\epsilon(r)$
not become negative if the denominator in Eq.~(\ref{eps}) is replaced
by $r\pm \epsilon(r)^{1/2}$. This implies $\epsilon(r) > r^{-2}$, and
limits the entropy of configurations as in example 1 to $S \sim
R^{5/2}$. This is still potentially problematic for the area entropy
of black holes. A limit of $S < A$ would require that $\epsilon(r) >
r^{-1}$. This would be the consequence of the previous logic if one
assumed a Planck length uncertainty in the radial coordinate $r$
rather than the proper radial distance $r \, \epsilon(r)^{-1/2}$ (or
equivalently an uncertainty in proper radial distance which grows as
$\epsilon(r)^{-1/2}$). This seems unphysical, but nevertheless cannot
be excluded as a consequence of quantum gravity. For related ideas,
see the stretched horizon in string theory \cite{Susskind}.

Below, we discuss the pathological properties of the 
configurations which exceed the $A^{3/4}$ bound.

\bigskip
{\bf Destroy all monsters!}
\bigskip

To obtain entropy scaling faster than $A^{3/4}$, we must consider
configurations in which $\epsilon (r)$ is close to zero in regions
containing significant entropy and energy density. We now show that
such configurations have the following pathological properties.

\smallskip
I. They inevitably evolve into black holes, even in the absence of any
   outside perturbation.

II. Even their {\it time-reversed} evolution leads to black hole
   formation.

\smallskip
They are therefore neither ordinary black holes nor ordinary matter
configurations. We refer to them as {\it monsters}.

To demonstrate I and II we consider the critical escape angle
$\theta_c \sim \epsilon (r)^{1/2}$ \cite{MTW}. Only particles whose
trajectories make an angle less than $\theta_c$ with the outward radial
direction can escape to infinity. All others follow orbits which bring
them to smaller $r$. (This phenomenon also contributes to the
persistence of a black hole atmosphere, or stretched horizon
\cite{membrane}.) A highly entropic configuration -- i.e., one in
which individual particle states have nearly randomly distributed
momenta -- with small $\epsilon (r)$ cannot avoid net energy flow
towards $r=0$ in its future evolution. This means that $\epsilon (r)$
will eventually cross zero and become negative, leading to horizon
formation. 

These conclusions
apply as well to the time-reversed evolution, since the
time-reversal of a highly entropic configuration is similar to the
original. (If momenta of individual particles in the configuration are
randomly distributed, then so are time-reversed momenta.)
A small subset of configurations, with entropy density reduced by a
factor of $\theta_c^2 \sim \epsilon(r)$, can avoid I and II if their
individual particle momenta are all nearly radial, or equivalently if
all modes are nearly S-wave. However, the reduction in entropy density
by $\epsilon (r)$ implies that the total entropy of such
configurations is less than that of flat space configurations.

In fact, if one defines a black hole as a region whose future does not
include future null infinity, then most of a monster configuration
already comprises a black hole \cite{fn1}. Even particles with exactly
radial trajectories cannot escape if they are deep inside -- the
infall of modes closer to the surface will cause a horizon to form
before they can escape. Roughly speaking, a configuration can have no
more than a mass fraction of order $\epsilon_0$ in a region with
$\epsilon(r) \sim \epsilon_0$ without eventually becoming a black
hole.

One might argue that it is impossible to create a monster, since it
turns into a black hole when evolved backwards in time: we would have
to begin with a {\it white hole}. However, the argument is not
conclusive: we could start with a normal configuration with the same
quantum numbers (e.g., ADM mass, charge, etc.) as the monster, which
tunnels or fluctuates quantum mechanically into the monster
state. There must be a nonzero, albeit very small, probability for
this if no conservation law is violated. Unless this process is
forbidden by new physics, it implies at least $\exp S$ black hole
microstates, where $S$ can grow faster than $A$ and may even be
unbounded. Note, though, that these states are inaccessible to
observers outside the hole. They cannot affect aspects of black hole
thermodynamics involving physics outside the horizon.

Clearly, monsters pose an interesting challenge to the interpretation
of black hole entropy as the logarithm of the number of microstates.
Nevertheless, the interpretation that $S = A/4$ represents the number
of ways to construct a hole out of ordinary matter and other
(non-monster) black holes still seems self-consistent, as we discuss
below.

\bigskip
{\bf Growing a black hole}
\bigskip

If we exclude monsters from consideration, ordinary matter
configurations have much less entropy than black holes of similar size
or mass. Almost all of the entropy of a given black hole must result
from a smaller black hole which has absorbed some additional mass.
This is the picture that has been developed in the membrane paradigm
\cite{membrane, FP} within a quasistationary
approximation.

Consider a black hole of area $A'$ that results from a hole of area
$A$ eating a small amount of energy $m$. We must have $\exp A' = \exp A \cdot 
\exp S$, where $S$ is the matter entropy. There must exist matter
configurations of mass $m$ near a black hole horizon which have
entropy $S$ of order $Mm$, since $A' - A \sim (M + m)^2 - M^2 \sim M
m$.

One can construct thin shell examples with mass $m$ and entropy $M m$, 
again taking advantage of curved space. Consider a
shell of thickness $d$ just outside the horizon, with energy density
 ($y = r-R$): $\rho(y) \sim m / d R^2$ and $\epsilon (r) \sim y/R$. 
Its entropy is
\begin{equation}
S \sim \int_0^d dy \, (R+y)^2 \, \left( \frac{R}{y} \right)^{1/2} 
\left( \frac{m}{d}\right)^{3/4} R^{-3/2} \sim \frac{M m}{(md)^{1/4}}~~.
\end{equation}
If one requires that the energy density $\rho$ be comprised of thermal
modes with wavelength $\lambda \sim \rho^{-1/4}$ less than
$\sqrt{Rd}$, the proper width of the shell, one obtains the constraint
that $md \sim 1$, so $S \sim M m$ as desired.

It is also worth noting that a single s-wave mode with energy $m \sim 1/R
\sim 1/M$ has entropy ${\cal O}(1)$, so satisfies $S \sim M m$. Thus,
a black hole can move along the $S \sim A$ curve by absorbing such
modes. This is arguably the smallest amount of energy that can be
absorbed by the hole, since otherwise the Compton wavelength of the
mode is much larger than the horizon itself.

\bigskip

{\bf Note added}: After this work was completed we were informed of related results obtained by Sorkin, Wald and Zhang \cite{Sorkin}. Those authors investigated monster-like objects as well as local extrema of the entropy $S$ subject to an energy constraint, which correspond to static configurations and obey $A^{3/4}$ scaling.
For example, in the case of a perfect fluid the local extrema satisfy the Tolman--Oppenheimer--Volkoff equation of hydrostatic equilibrium. In considering monster configurations, Sorkin et al. show that requiring a configuration to be no closer than a thermal wavelength $\lambda \sim \rho^{-1/4}$ from its Schwarzschild radius imposes the bound $S < A$. While this may be a reasonable criterion that must be satisfied for the assembly of an initial configuration, it does not seem to apply to states reached by quantum tunneling. Note that from a global perspective configurations with $S > A^{3/4}$ are already black holes in the sense that the future of the interior of the object does not include future null infinity.

\bigskip

\emph{Acknowledgements---} The authors thank R. Buniy, S. Carlip, M. Einhorn, R. Sorkin and  
A. Zee for useful comments.
The authors are supported by the
Department of Energy under DE-FG02-96ER40969.

%\newpage

%%%%%%%%%%%%%%%%%%%%%%%%%%%%%%%%%%%%%%%%%%%%%%%%%%%%%%%%%%%%%%%%%
%%%
%%%                     BIBLIOGRAPHY
%%%
%%%%%%%%%%%%%%%%%%%%%%%%%%%%%%%%%%%%%%%%%%%%%%%%%%%%%%%%%%%%%%%%%

\bigskip

%\newpage
%\vskip .75 in


\begin{thebibliography}{99}
\bibitem{fn} A collection of $N>1$ smaller black holes with masses $m=M/N$ only yields a subleading contribution $Nm^2 \sim A/N$ to the entropy unless curved space configurations similar to the monsters discussed later are considered.
\end{thebibliography}
\end{document}